\documentclass[12pt]{article}
\usepackage{graphicx}
\usepackage{epstopdf}
\usepackage{amssymb}
\usepackage{amsmath}
\begin{document}
\baselineskip 0.6cm
\newcommand{\tri}{\triangleright}
\newcommand{\range}{{\rm range}}
\newcommand{\Ree}{{\rm Re }}
\newcommand{\Imm}{{\rm Im }}
\newcommand{\diag}{{\rm diag}}
\newcommand{\sign}{{\rm sign}}
\newcommand{\tr}{{\rm tr}}
\newcommand{\rank}{{\rm rank}}
\newcommand{\bp}{\bigskip}
\newcommand{\mdp}{\medskip}
\newcommand{\slp}{\smallskip}
\newcommand{\Rw}{\Rightarrow}
\newcommand{\ts}{& \hspace{-0.1in}}
\newcommand{\nn}{\nonumber}
\newcommand{\bea}{\begin{eqnarray}}
\newcommand{\eea}{\end{eqnarray}}
\newcommand{\beas}{\begin{eqnarray*}}
\newcommand{\eeas}{\end{eqnarray*}}
\newcommand{\beq}{\begin{equation}}
\newcommand{\eeq}{\end{equation}}
\newtheorem{exa}{Example}[section]
\newtheorem{thm}{Theorem}[section]
\newtheorem{lem}{Lemma}[section]
\newtheorem{prop}{Proposition}[section]
\newtheorem{fact}{Fact}[section]
\newtheorem{cor}{Corollary}[section]
\newtheorem{defn}{Definition}[section]
\newtheorem{rem}{Remark}[section]
\renewcommand{\theequation}{\thesection.\arabic{equation}}

\title{{\bf Adaptive Output Tracking Control with Reference Model
    System Uncertainties: Extensions}}
\author{{\it Gang Tao}\\
\normalsize Department  of Electrical and Computer Engineering\\  
\normalsize University of Virginia \\ 
\normalsize Charlottesville, VA 22903, USA} 
\date{} 
\maketitle 
\begin{abstract}
This paper develops some extensions to the work of \cite{t24b} which
studied the continuous-time adaptive output tracking control schemes
 with the reference output signal generated from an
unknown reference model system. The presented extensions include
adaptive control schemes with reference model system uncertainties for
single-input single-output (SISO) discrete-time systems and
multi-input multi-output (MIMO) discrete-time, continuous-time and
feedback linearizable systems as well. To deal with such reference
model system uncertainties, the adaptive controller structures are
expanded to include a parametrized estimator of the equivalent
reference input signal, to ensure a completely parametrized error system with a
known regressor vector, suitable for stable adaptive controller
parameter update law design.
\end{abstract}

\bigskip
\begin{quote}
\end{quote}

\setcounter{equation}{0}
\section{Introduction}
In \cite{t24b}, we considered a linear time-invariant (LTI) system (plant) described by
\beq
\dot{x}(t) = A x(t) + b u(t),\;y(t) = c x(t),
\label{5x1}
\eeq
where $x(t) \in R^{n}$, $u(t) \in R$ and $y(t) \in R$ are the state
vector, input and output signals, and $A \in R^{n \times n}$, $b \in
R^{n}$ and $c \in R^{1 \times n}$ are unknown constant matrices such
that $G(s) = c(sI - A)^{-1}b$ has relative degree $n^*$. 
The control objective is to design the input signal $u(t)$
to ensure the closed-loop system stability and asymptotic output
tracking: $\lim_{t \rightarrow \infty}(y(t) - y_m(t)) = 0$, where
$y_m(t)$ is the output of a stable reference model system:
\beq
\dot{x}_{m}(t) = A_{m} x_{m}(t) + b_{m} u_m(t),\;y_m(t) = c_m x_m(t)
\label{5x0}
\eeq
where $x_{m}(t) \in R^{n}$, $u_m(t) \in R$ and $y_m(t) \in R$ are
the reference system state vector, input and output signals, and
$A_{m} \in R^{n \times n}$, $b_{m} \in R^{n}$ and $c_m \in R^{1 \times
  n}$ are some constant matrices such that $G_m(s) = c_m(sI -
A_m)^{-1} b_m$ has relative degree $n_m^* \geq n^*$.

\medskip
The output tracking problem in \cite{t24b} was different from the
traditional model reference adaptive control problem in which 
 the reference model system has an input-output system form
\beq
y_m(t) = W_m(s)[r_m](t),
\label{5mrs}
\eeq
where $W_m(s) = \frac{1}{P_m(s)}$ for a chosen stable polynomial
$P_m(s)$ of degree $n^*$ (as a convenient notation, $y_m(t) = W_m(s)[r_m](t)$
denotes the output $y_m(t)$ of a system $W_m(s)$ with input
$r_m(t)$), and $r_m(t)$ is a given (available) reference input. Both 
$W_m(s)$ and $r_m(t)$ are used in a MRAC scheme.

In terms of $G_m(s)$ and $u_m(s)$, we have $y_m(t) =
G_m(s)[u_m](t) = W_m(s) P_m(s) G_m(s)[u_m](t)$, so that $r_m(t) = P_m(s)
G_m(s)[u_m](t)$ which is known and available from $u_m(t)$ if $G_m(s)$
is known. The adaptive output tracking problem solved in \cite{t24b}
is for the case when the reference model system $(A_m, b_m,
c_m)$ or $G_m(s) = c_m(sI - A_m)^{-1} b_m$ is unknown and 
the time-derivatives $y_m^{(i)}(t)$, $i=1,2,\ldots, n^*$, are unknown
either (such derivative information is also sufficient for MRAC). 

\medskip
The solutions to such a new MRAC problem, developed in \cite{t24b},
were for SISO continuous-time LTI systems. In this paper, we extend
the results of \cite{t24b} to SISO discrete-time LTI systems (in
Section 2), to MIMO LTI systems (in Section 3, with a unified
presentation for both discrete-time and continuous-time systems and
adaptive control designs), and to MIMO feedback linearizable nonlinear
systems (in Section 4). More details of the compact presentations
given in this paper about such extensions can be found in \cite{t24}.

\setcounter{equation}{0}
\section{Discrete-Time Designs for SISO Systems}
The problem can be formulated for a SISO discrete-time LTI plant
\beq
x(t+1) = A x(t) + b u(t),\;y(t) = c x(t),
\label{5x1d}
\eeq
where $x(t) \in R^{n}$, $u(t) \in R$ and $y(t) \in R$ are
plant state vector, input and output signals, and $A \in R^{n \times 
n}$, $b \in R^{n}$ and $c \in R^{1 \times n}$ are unknown, and so is
the plant transfer function $G(z) = c(zI - A)^{-1}b$ which has
a known relative degree $n^*$ and all its zeros stable.

\medskip
The control objective is to design a feedback control signal
$u(t)$ to ensure closed-loop signal boundedness and asymptotic
$y(t)$ tracking the output $y_m(t)$ of a reference model system 
\beq
x_{m}(t+1) = A_{m} x_{m}(t) + b_{m} u_m(t),\;y_m(t) = c_m x_m(t)
\label{5x0d}
\eeq
where $u_m(t) \in R$ and $y_m(t) \in R$ (and $x_{m}(t) \in R^{n}$) are
known, $A_{m} \in R^{n \times n}$, $b_{m} \in R^{n}$ and
$c_m \in R^{1 \times n}$ are unknown constant matrices, and so is the
reference system transfer function $G_m(z) = c_m(z I - A)^{-1} b_m$
which is assumed to have a known relative degree $n_m^* \geq n^*$.

\medskip
The main adaptive control design idea is similar to that for the continuous-time
case \cite{t24b}, but in a discrete-time framework: for a chosen stable
polynomial $P_m(z)$ of degree $n^*$: 
\beq
P_m(z) = z^{n^*} + p_{n^*-1} z^{n^*-1}
+ \cdots + p_1 z + p_0, 
\eeq
we construct an estimate of the equivalent reference model system input
\bea
r_m(t) \ts =\ts  P_m(z)[y_m](t) \nn\\
\ts = \ts y_m(t+n^*) + p_{n^*-1} y_m(t+n^*-1) + \cdots + p_1 y_m(t+1) + p_0 y_m(t),
\label{rm(t)d}
\eea
such that $y_m(t) = G_m(z)[u_m](t) = \frac{1}{P_m(z)}[r_m](t)$, that is,
$
r_m(t) = P_m(z) G_m(z)[u_m](t).
$

If the signals $y_m(t+i)$, $i=1,2,\ldots, n^*$, were known, or
$G_m(s)$ were known, we could obtain $r_m(t)$ from either 
$r_m(t) = P_m(z)[y_m](t)$ or $r_m(t) = P_m(z) G_m(z)[u_m](t)$, for
adaptive control design. But, they are unknown in the current control
problem, so that $r_m(t)$ needs to be estimated whose estimate is the
estimate of $r_m(t) = P_m(z)[y_m](t)$ in a parametrized form.

\subsection{State Feedback Control Designs}
\label{State Feedback Control Designs}
We first consider adaptive state feedback control designs using either
the reference model system signals $u_m(t)$ and $x_m(t)$ or the
signals $u_m(t)$ and $y_m(t)$.

\subsubsection{Design with $x_m(t)$ Available}
We derive a parametrized nominal state feedback controller
structure for ($A$, $b$, $c$) and ($A_m$, $b_m$, $c_m$) known, and 
then develop its adaptive version for ($A$, $b$, $c$) and ($A_m$,
$b_m$, $c_m$) unknown.

\medskip
The condition that the reference system (\ref{5x0d}) with relative degree
$n_m^*$, implies: $c_m A_m^i b_m = 0$ for $i= 0,1,\ldots, n_m^*-2$,
$c_m A_m^{n_m^*-1} b_m \not = 0$, so that
\beq
y_m(t+i) = \left\{ \begin{array}{ll}
c_m A_m^{i} x_m(t) & \mbox{for $i=0,1,\ldots,n_m^* - 1$} \\ 
c_m A_m^{i} x_m(t) + c_m A_m^{i-1} b_m u_m(t) & \mbox{for $i=n_m^*$.}
\end{array}
\right.
\eeq
Hence, for $n^* \leq n_m^*$, we can express $r_m(t)$ in (\ref{rm(t)d}) as
\beq
r_m(t) = \alpha_1^T x_m(t) + \alpha_2 u_m(t)
\label{5rm(t)d}
\eeq
for some parameters $\alpha_1 \in R^n$ and $\alpha_2 \in R$, related
to the reference system parameters ($A_m$, $b_m$, $c_m$).

\bigskip
{\bf Nominal control law}. The nominal state feedback control law is
\bea
u(t) \ts = \ts k_1^{*T} x(t) + k_2^* r_m(t) \nn \\
\ts = \ts  k_1^{*T} x(t) + 
k_{21}^{*T} x_m(t) + k_{22}^* u_m(t),
\label{5u1nd}
\eea
where 
\beq
k_{21}^{*} = k_2^* \alpha_1,\;k_{22}^* = k_2^* \alpha_2.
\eeq
and $k_1^{*} \in R^n$ and $k_2^* \in R$ are constant
parameters satisfying
\beq
c(z I - A - b k_1^{*T})^{-1} b k_2^* = W_m(z) = \frac{1}{P_m(z)}.
\eeq
The output of the closed-loop control system is
\beq
y(t) = c(zI - A - b k_1^{*T})^{-1} b k_2^*[r_m](t)
= W_m(z)[r_m](t) = y_m(t), 
\eeq
with the exponentially decaying initial condition effect ignored.

\bigskip
{\bf Adaptive control law}. The adaptive control law is
chosen as
\bea
u(t) \ts = \ts  k_1^{T} x(t) + 
k_{21}^{T} x_m(t) + k_{22} u_m(t),
\label{5u1ad}
\eea
where $k_1 \in R^n$, $k_{21} \in R^n$ and $k_{22} \in R$ are the
estimates of $k_1^*$, $k_{21}^*$ and $k_{22}^*$.

\bigskip
{\bf Error model}. With (\ref{5u1ad}), the closed-loop system with
(\ref{5x1d}) is 
\bea
x(t+1) \ts = \ts A x(t) + b (k_1^{*T} x(t) + 
k_{21}^{*T} x_m(t) + k_{22}^* u_m(t)) \nn\\
\ts \ts + b ((k_1- k_1^*)^{T} x(t) + 
(k_{21}- k_{21}^*)^{T} x_m(t) + (k_{22}-k_{22}^*) u_m(t)).\;\;\;\;\;\;\;\;\;
\label{5x1cd}
\eea
With $k_{21}^{*T} x_m(t) + k_{22}^* u_m(t) = k_2^* r_m(t)$ and 
$y_m(t) = c(zI - A - b k_1^{*T})^{-1} b k_2^*[r_m](t) =
W_m(z)[r_m](t) = G_m(z)[u_m](t)$ known, for $\rho^* = \frac{1}{k_2^*}$ and 
the tracking error $e(t) = y(t)- y_m(t)$ with $y(t) = c x(t)$, from
(\ref{5x1cd}), we derive the tracking error equation
\beq
e(t) = \rho^* W_m(z)[(k_1- k_1^*)^{T} x + 
(k_{21}- k_{21}^*)^{T} x_m + (k_{22}-k_{22}^*) u_m](t).
\label{5e(t)d}
\eeq

\medskip
We introduce the estimation error signal
\beq
\epsilon (t) =  e(t) + \rho(t) \xi(t),
\label{5732A.35d}
\eeq
where $\rho(t)$ is the estimate of $\rho^{*}$, and 
\bea
\xi(t) \ts = \ts \theta^{T}(t) \zeta(t) - W_m (z) [\theta^{T} \omega](t) \in R\\
\theta (t) \ts = \ts \left[k_1^T(t), k_{21}^T(t), k_{22}(t)\right]^T \in R^{2n+1}\\
\omega (t) \ts = \ts \left[x^{T}(t), x_m^T(t), u_m(t) \right]^{T}\in R^{2n+1}\\
\zeta (t) \ts = \ts W_{m}(z)[\omega](t)\in R^{2n+1}.
\eea

For $\theta^{*} = \left[k_1^{*T}, k_{21}^{*T}, k_{22}^* \right]^T\in
R^{2n+1}$, it can be verified that
\beq
\epsilon (t) = \rho^{*} (\theta(t) - \theta^{*})^{T} \zeta(t) + (\rho(t) - 
\rho^{*}) \xi(t),
\label{5732A.36d}
\eeq
which is a completely parametrized linear model, in terms of the
transformed plant parameters and reference model system parameters in
$\theta^*$ and their estimates in $\theta(t)$.

\medskip
{\bf Adaptive laws}. 
The adaptive laws for $\theta(t)$ and $\rho(t)$ are chosen as
\bea
\label{5732A.37d}
\theta(t+1) \ts = \ts \theta(t) - \frac{\Gamma \sign[k_p] \zeta (t) \epsilon (t)} 
{m^2(t)} \\*[0.05in]
\rho(t+1) \ts = \ts \rho(t) - \frac{\gamma\, \xi(t) \epsilon (t)}{m^2(t)}
\label{5732A.38d}
\eea
where $0< \Gamma = \Gamma^T < \frac{2}{|k_p|} I$ and $0 < \gamma < 2$ are adaptation gains, 
and 
\beq
m(t) = \sqrt{1 + \zeta^{T}(t) \zeta(t) + \xi^{2}(t)}.
\label{5732A.39d}
\eeq

\medskip
This adaptive scheme has the following desired and standard adaptive
law properties.

\begin{lem} 
\label{5lemma11d}
The adaptive laws (\ref{5732A.37d}) and (\ref{5732A.38d}) ensure that 
$\theta(t) \in L^{\infty}$, $\rho(t) \in L^{\infty}$, 
 $\frac{\epsilon(t)}{m(t)} \in L^{2} \cap L^{\infty}$,
 $\theta(t+1)-\theta(t) \in L^{2}$, and 
 $\rho(t+1) - \rho(t) \in L^{2}$.
\end{lem} 

\subsubsection{Design with $x_m(t)$ Unavailable}
Consider the reference model system (\ref{5x0d}): 
\beq
x_{m}(t+1) = A_{m} x_{m}(t) + b_{m} u_m(t),\;y_m(t) = c_m
x_m(t),
\eeq
with $x_m(t)$ unavailable. We can use a nominal reduced-order state
observer \cite[page 553]{r96}, \cite[page 190]{t03}, based on the
knowledge of $y_m(t)$ and $u_m(t)$, to
generate an estimate $\hat{x}_{m}(t)$ of $x_m(t)$ such that 
$\lim_{t \rightarrow \infty} (\hat{x}_{m}(t) - x_m(t)) = 0$
exponentially. Then, replacing $x_m(t)$ with $\hat{x}_m(t)$, we can
express $r_m(t) = \alpha_1^T \hat{x}_m(t) + \alpha_2 u_m(t)$ as
\beq
r_m(t) = \beta_1^T \omega_{u_m}(t) + \beta_{2}^T \omega_{y_m}(t) +
\beta_{20} y_m(t) + \alpha_2 u_m(t)
\label{5r02d}
\eeq
for some $\beta_1 \in R^{n-1}$, $\beta_{2} \in R^{n-1}$ and
$\beta_{20} \in R$, where
\beq
\omega_{u_m}(t) = F(z)[u_m](t),\;\omega_{y_m}(t) = F(z)[y_m](t),
\eeq
\beq
F(z) = \frac{a(z)}{\Lambda_e(z)},\;a(z) = [1, z, \ldots, z^{n - 2}]^{T},
\eeq
for a chosen monic stable polynomial $\Lambda_e(z)$ of degree $n-1$.

\bigskip
{\bf Nominal control law}. The nominal state feedback control law is
\bea
u(t) \ts = \ts k_1^{*T} x(t) + k_2^* r_m(t) \nn \\
\ts = \ts  k_1^{*T} x(t) + k_{21}^{*T} \omega_{u_m}(t) +
k_{22}^{*T} \omega_{y_m}(t) + k_{20}^* y_m(t) + k_3^* u_m(t),
\label{5u1n1d}
\eea
where 
\beq
k_{21}^{*} = k_2^* \beta_1,\;k_{22}^* = k_2^* \beta_2,\;k_{20}^* =
k_2^* \beta_{20},\;k_3^* = k_2^* \alpha_2.
\eeq

\medskip
{\bf Adaptive control law}. The adaptive state feedback control law is
\bea
u(t)\ts =\ts  k_1^{T} x(t) + k_{21}^{T} \omega_{u_m}(t) +
k_{22}^{T} \omega_{y_m}(t) + k_{20} y_m(t) + k_3 u_m(t)\nn\\
\ts = \ts \theta^T(t) \omega(t),
\label{5u1n1ad}
\eea
where $k_1$, $k_{21}$, $k_{22}$, $k_{20}$ and $k_3$ are the estimates
of $k_1^*$, $k_{21}^{*}$, $k_{22}^*$, $k_{20}^*$ and $k_3^*$, and
\bea
\theta(t) \ts = \ts [k_1^T(t), k_{21}^T(t), k_{22}^T(t), k_{20}(t),
k_3(t)]^T \in R^{3n}\\
\omega(t) \ts = \ts [x^T(t), \omega_{u_m}^T(t), \omega_{y_m}^T(t),
y_m(t), u_m(t)]^T \in R^{3n}.
\eea

\medskip
{\bf Error model}. With the control law (\ref{5u1n1ad}), the
closed-loop tracking error equation for $e(t) = y(t) - y_m(t)$ can be
derived as
\beq
e(t) = \rho^* W_m(z)[(\theta - \theta^*)^T \omega](t), 
\label{5e(t)1d}
\eeq
which has the same form as that in (\ref{5e(t)d}), where
\beq
\theta^* = [k_1^{*T}, k_{21}^{*T}, k_{22}^{*T}, k_{20}^*,
k_3^*]^T \in R^{3n}.
\eeq

Similarly, based on (\ref{5e(t)1d}), an estimation error
$\epsilon(t)$ can be introduced:
\beq
\epsilon (t) =  e(t) + \rho(t) \xi(t),
\eeq
where $\rho(t)$ is the estimate of $\rho^{*}$, and 
\bea
\xi(t) \ts = \ts \theta^{T}(t) \zeta(t) - W_m (z) [\theta^{T} \omega](t) \in R\\
\zeta (t) \ts = \ts W_{m}(z)[\omega](t)\in R^{3n},
\eea
and the adaptive laws for $\theta(t)$ and $\rho(t)$ can be chosen as
\bea
\theta(t+1) \ts = \ts \theta(t) - \frac{\Gamma \sign[k_p] \zeta (t) \epsilon (t)} 
{m^2(t)} \\*[0.05in]
\rho(t+1) \ts = \ts \rho(t) - \frac{\gamma\, \xi(t) \epsilon (t)}{m^2(t)}
\eea
where $0< \Gamma = \Gamma^T < \frac{2}{|k_p|} I$ and $0 < \gamma < 2$, and 
$m(t) = \sqrt{1 + \zeta^{T}(t) \zeta(t) + \xi^{2}(t)}$, to ensure
the desired properties similar to that given in Lemma \ref{5lemma11d}.

\subsection{Output Feedback Control Designs}
An output feedback control law can be designed for the 
discrete-time plant (\ref{5x1d}): $x(t+1) = A x(t) + b u(t),\;y(t) = c
x(t)$, with the reference model system (\ref{5x0d}): 
$x_{m}(t+1) = A_{m} x_{m}(t) + b_{m} u_m(t),\;y_m(t) = c_m x_m(t)$,
starting with the traditional output feedback controller structure
\begin{equation}
u(t) = \theta_{1}^{T}\omega_{1}(t) + \theta_{2}^{T}\omega_{2}(t) + \theta_{20}
 y(t) + \theta_{3} r_m(t),
\label{5732A.93}
\end{equation}
where $r_m(t) = P_m(z) G_m(z)[u_m](t) = P_m(z)[y_m](t)$ with $P_m(z)$ being a chosen
stable polynomial of degree $n^*$ equal to the relative degree of
$G(z) = c(z I - A)^{-1} b$ and $G_m(z) = c_m(z I - A_m)^{-1} b_m$
whose relative degree is $n_m^* \geq n^*$, 
$\theta_{1} \in R^{n-1}$, $\theta_{2} \in R^{n-1}$, $\theta_{20}
\in R$ and $\theta_{3} \in R$, and 
\begin{equation}
\omega_{1}(t) = \frac{a(z)}{\Lambda(z)}[u](t),\;\omega_{2}(t) = 
\frac{a(z)}{\Lambda(z)}[y](t)
\end{equation}
with $a(z) = [1,z,\cdots,z^{n-2}]^{T}$ and $\Lambda(z)$ being a monic
stable polynomial of degree $n-1$. 
 The parameters $\theta_{1} \in R^{n-1}$, $\theta_{2} \in R^{n-1}$, $\theta_{20}
\in R$ and $\theta_{3} \in R$ are the estimates of the constant parameters
$\theta_{1}^{*} \in R^{n-1}$, $\theta_{2}^{*} \in R^{n-1}$, 
$\theta_{20}^{*} \in R$  and $\theta_3^* = \frac{1}{k_p}$, which satisfy the polynomial matching equation:
\beq
 \theta_{1}^{*T} a(z) P(z) 
+ (\theta_{2}^{*T} a(z) + \theta_{20}^{*} 
\Lambda(z)) k_{p} Z(z) = \Lambda(z)(P(z) - k_{p}  
\theta_{3}^{*} Z(z) P_{m}(z)),
\label{5732A.95}
\eeq
for $G(z) = c(zI - A)^{-1} b = k_p \frac{Z(z)}{P(z)}$ with monic
polynomials $Z(z)$ and $P(z)$ of degrees $n-n^*$ and $n$.

\bigskip
{\bf Parametrization of $r_m(t)$}. 
In the current output tracking control problem, 
the reference model system ($A_m$, $b_m$, $c_m$) or $G_m(s)$ is unknown,
we cannot directly generate $r_m(t)$ from $r_m(t) = P_m(z)
G_m(z)[u_m](t)$ for a traditional adaptive controller. In
stead, we use the equivalent expression $r_m(t) = P_m(z)[y_m](t)$ to
obtain an estimate of $r_m(t)$, to be embedded in the control law.

\medskip
As derived in (\ref{5rm(t)d}), $r_m(t)$ has the expression
\beq
r_m(t) = \alpha_1^T x_m(t) + \alpha_2 u_m(t)
\label{5rm(t)u}
\eeq
for some parameters $\alpha_1 \in R^n$ and $\alpha_2 \in R$, related
to the reference system parameters ($A_m$, $b_m$, $c_m$).

\medskip
When $x_m(t)$ is not available, using (\ref{5r02d}), we have
\beq
r_m(t) = \beta_1^T \omega_{u_m}(t) + \beta_{2}^T \omega_{y_m}(t) +
\beta_{20} y_m(t) + \alpha_2 u_m(t)
\label{5r02u}
\eeq
for some $\beta_1 \in R^{n-1}$, $\beta_{2} \in R^{n-1}$ and
$\beta_{20} \in R$, where
\beq
\omega_{u_m}(t) = F(z)[u_m](t),\;\omega_{y_m}(t) = F(z)[y_m](t),
\eeq
\beq
F(z) = \frac{a(z)}{\Lambda_e(z)},\;a(z) = [1, z, \ldots, z^{n - 2}]^{T},
\eeq
for a chosen monic stable polynomial $\Lambda_e(z)$ of degree $n-1$.

\subsubsection{Nominal Control Laws} 
A nominal control law can be designed for $x_m(t)$ available or
unavailable, based on the parameters of the  plant and the reference
model system, whose structure is used for adaptive control design.

\bp
{\bf Nominal control law for $x_m(t)$ available}. 
With $x_m(t)$ available, the nominal control law is
\bea
u(t) \ts = \ts \theta_{1}^{*T}\omega_{1}(t) + \theta_{2}^{*T}\omega_{2}(t) + \theta_{20}^*
 y(t) + \theta_{3}^* r_m(t) \nn\\
\ts = \ts \theta_{1}^{*T}\omega_{1}(t) + \theta_{2}^{*T}\omega_{2}(t) + \theta_{20}^*
 y(t) + \theta_{3}^* (\alpha_1^T x_m(t) + \alpha_2 u_m(t)) \nn\\
\ts = \ts \theta_{1}^{*T}\omega_{1}(t) + \theta_{2}^{*T}\omega_{2}(t) + \theta_{20}^*
 y(t) + \theta_{31}^{*T} x_m(t) + \theta_{32}^* u_m(t) \nn\\
\ts = \ts \theta^{*T}\omega(t), 
\label{5732A.931}
\eea
where 
\beq
\theta_{31}^* = \theta_3^* \alpha_1 \in R^n,\;
\theta_{32}^* = \theta_3^* \alpha_2 \in R
\eeq
\beq
\theta^* =
[\theta_1^{*T}, \theta_2^{*T}, \theta_{20}^*, \theta_{31}^{*T}, \theta_{32}^*]^T \in R^{3n}
\label{5thetasa}
\eeq
\beq
\omega(t) = [\omega_1^T(t), \theta_2^T(t), y(t), x_m^T(t),
u_m(t)]^T \in R^{3n}.
\label{5omegaa}
\eeq

Such a nominal control law can achieve the desired control objective:
all closed-loop system signals are bounded and
$\lim_{t \rightarrow \infty} (y(t) - y_m(t)) = 0$
(exponentially). 

\bp
{\bf Nominal control law for $x_m(t)$ unavailable}. 
When $x_m(t)$ is not available, we use (\ref{5r02u}) to construct the
control law
\bea
u(t) \ts = \ts \theta_{1}^{*T}\omega_{1}(t) + \theta_{2}^{*T}\omega_{2}(t) + \theta_{20}^*
 y(t) + \theta_{3}^* r_m(t) \nn\\
\ts = \ts \theta_{1}^{*T}\omega_{1}(t) + \theta_{2}^{*T}\omega_{2}(t) + \theta_{20}^*
 y(t) \nn\\
\ts \ts + \theta_{3}^* (\beta_1^T \omega_{u_m}(t) + \beta_{2}^T \omega_{y_m}(t) +
\beta_{20} y_m(t) + \alpha_2 u_m(t)) \nn\\
\ts = \ts \theta_{1}^{*T}\omega_{1}(t) + \theta_{2}^{*T}\omega_{2}(t) + \theta_{20}^*
 y(t) \nn\\
\ts \ts +  \theta_{31}^{*T} \omega_{u_m}(t) + \theta_{32}^{*T} \omega_{y_m}(t) +
\theta_{33}^* y_m(t) + \theta_{34}^* u_m(t)\nn\\
\ts = \ts \theta^{*T}\omega(t), 
\label{5732A.932}
\eea
where
\beq
\theta_{31}^* = \theta_{3}^* \beta_1,\;
\theta_{32}^* = \theta_{3}^* \beta_2,\;
\theta_{33}^* = \theta_{3}^* \beta_{20},\;
\theta_{34}^* = \theta_{3}^* \alpha_2
\eeq
\beq
\theta^* =
[\theta_{1}^{*T}, \theta_{2}^{*T}, \theta_{20}^*, \theta_{31}^{*T}, \theta_{32}^{*T}, \theta_{33}^*, \theta_{34}^*]^T \in R^{4n-1}
\label{5thetasu} 
\eeq
\beq
\omega(t) = [\omega_{1}^T(t), \omega_{2}^T(t), 
y(t), \omega_{u_m}^T(t), \omega_{y_m}^T(t), y_m(t), u_m(t)]^T.
\label{5omegau}
\eeq

\subsubsection{Adaptive Control Schemes} 
Adaptive control laws use the nominal controller structures with parameter estimates.

\bp
{\bf Adaptive control law for $x_m(t)$ available}.
Based on the nominal control law (\ref{5732A.931}), the adaptive
control law is
\beq
u(t) = \theta^T(t) \omega(t),
\label{5uu(t)1}
\eeq
where $\omega(t)$ is defined in (\ref{5omegaa}), and $\theta(t)$ is
the estimate of $\theta^*$ defined in (\ref{5thetasa}). 

\bp
{\bf Adaptive control law for $x_m(t)$ unavailable}. 
Based on the nominal control law (\ref{5732A.932}), the adaptive
control law is
\beq
u(t) = \theta^T(t) \omega(t),
\label{5uu(t)2}
\eeq
where $\omega(t)$ is defined in (\ref{5omegau}), and $\theta(t)$ is
the estimate of $\theta^*$ defined in (\ref{5thetasu}). 

\bp
{\bf Tracking error equation}. 
In both cases, the tracking error $e(t) = y(t) - y_m(t)$ satisfies
\beq
e(t) = \rho^* W_m(z)[(\theta - \theta^*)^T \omega](t),
\label{5utee}
\eeq
with different $\theta$, $\theta^*$ and $\omega(t)$, which has the same
form as (\ref{5e(t)d}) or (\ref{5e(t)1d}), and can be used to define
the estimation error for adaptive laws to update $\theta(t)$ and the
estimate $\rho(t)$ of $\rho^*$.

\bp
{\bf Adaptive laws}. 
Based on (\ref{5utee}), we can similarly define the estimation error
\beq
\epsilon (t) =  e(t) + \rho(t) \xi(t),
\eeq
where $\rho(t)$ is the estimate of $\rho^{*}$, and 
\bea
\xi(t) \ts = \ts \theta^{T}(t) \zeta(t) - W_m (z) [\theta^{T} \omega](t) \in R\\
\zeta (t) \ts = \ts W_{m}(z)[\omega](t) \in R^{n_\theta}, 
\eea
and choose the adaptive laws for $\theta(t)$ and $\rho(t)$ as
\bea
\label{5uthetalaw}
\theta(t+1) \ts = \ts \theta(t) - \frac{\Gamma \sign[k_p] \zeta (t) \epsilon (t)} 
{m^2(t)} \\*[0.05in]
\rho(t+1) \ts = \ts \rho(t) - \frac{\gamma\, \xi(t) \epsilon (t)}{m^2(t)},
\label{5urholaw}
\eea
where $0< \Gamma = \Gamma^T < \frac{2}{|k_p|} I$ and $0 < \gamma < 2$,
and \beq
m(t) = \sqrt{1 + \zeta^{T}(t) \zeta(t) + \xi^{2}(t)}.
\eeq

\bigskip
{\bf Adaptive system properties}. The estimation error $\epsilon(t)$
satisfies
\beq
\epsilon (t) = \rho^{*} \tilde{\theta}^T(t) \zeta(t)
+ \tilde{\rho}(t) \xi(t),\;
\tilde{\theta}(t) = \theta(t) - \theta^{*},\tilde{\rho}(t) = \rho(t) - \rho^{*}.
\eeq
The time-increment of the positive definite function
$
V(\tilde{\theta}, \tilde{\rho}) = |\rho^{*}|\tilde{\theta}^{T} 
\Gamma^{-1} \tilde{\theta} + \gamma^{-1} \tilde{\rho}^{2},
$ is
\bea
\ts \ts V(\tilde{\theta}(t+1), \tilde{\rho}(t+1)) -
V(\tilde{\theta}(t), \tilde{\rho}(t)) \nn\\
\ts = \ts - \left(2 - \frac{|k_p| \zeta^T(t) \Gamma \zeta(t)
+ \gamma \xi^2(t)}{m^2(t)}\right) \frac{\epsilon^2(t)}{m^2(t)} \leq 0.
\eea

Then, we have the desired adaptive law properties given in Lemma
\ref{5lemma11d}, based on which we have:

\begin{thm}
\label{5thm1}
The adaptive controller (\ref{5uu(t)1}) or (\ref{5uu(t)2}), updated from the adaptive law
(\ref{5uthetalaw})-(\ref{5urholaw}) and applied to the plant
(\ref{5x1d}), ensures that all closed-loop system signals are bounded
and $\lim_{t \rightarrow \infty} (y(t) - y_m(t)) = 0$.
\end{thm}

The results of this theorem also hold for adaptive state feedback
control designs developed in Section \ref{State Feedback Control
  Designs}. The proofs of these results can be derived in a similar
way to that for a traditional model reference adaptive control scheme,
as the signal terms related to the reference model system are bounded,
and the $L^2$ properties of the adaptive laws hold.

\setcounter{equation}{0}
\section{Unified Designs for MIMO Systems}
Consider a multi-input multi-output linear time-invariant plant
\beq
\dot{x}(t) = A x(t) + B u(t),\;y(t) = C x(t),
\label{52eqn1}
\eeq
where $x(t) \in R^{n}$ and $y(t) \in R^M$ are the plant state and output
vectors, $u(t) \in R^{M}$ is the input vector, and $(A, B, C)$ are
unknown matrices.

The control objective is to design a feedback control input $u(t)$ to ensure
closed-loop system signal boundedness and asymptotic $y(t)$ tracking
the output $y_m(t)$ of a given reference system 
\beq
\dot{x}_m(t) = A_m x_m(t) + B_m u_m(t),\;y_m(t) = C_m x_m(t),
\label{52eqn1m}
\eeq
where $x_m(t) \in R^{n}$ and $y_m(t) \in R^M$ are the reference system
 state and output vectors, $u_m(t) \in R^{M}$ is a given input
 vector, and, unlike the traditional model reference adaptive control
 problem, $(A_m, B_m, C_m)$ are unknown parameter matrices, so
 that $G_m(s) = C_m(sI - A_m)^{-1} B_m$ is unknown.

\medskip
For adaptive control of a multivariable plant (\ref{52eqn1}) whose
transfer matrix is $G(s) = C (s I - A)^{-1} B$, its
modified interactor matrix $\xi_m(s)$ \cite[page 385]{t03} (which is a lower
triangular $M \times M$ polynomial matrix such that $\lim_{s
  \rightarrow \infty} \xi_m(s) G(s) = K_p$ is nonsingular and finite,
and $\xi_m^{-1}(s)$ is a stable rational matrix) is an important
structure information to use, as described in the following assumption:

\begin{description}
\item[] {\bf Assumption (A3.1)}: The modified interactor matrix
$\xi_m(s)$ of the plant $G(s)$ is known, and, for the reference model
  system $G_m(s)$,  $\lim_{s \rightarrow
  \infty} \xi_m(s) G_m(s)$ is finite.
\end{description}

For stable output tracking control, we also need the assumption:

\begin{description}
\item[] {\bf Assumption (A3.2)}: The plant $(A, B, C)$ is stabilizable
  and detectable, and all zeros of $G(s)$ are stable.
\end{description}

The above formulation also holds for the discrete-time case, with the symbol $s$
(which denotes either the time-differentiation operator: $s[x](t) =
\dot{x}(t)$ or the Laplace transform variable in the continuous-time
case) replaced by the symbol $z$ (which denotes either the
time-advance operator: $z[x](t) = x(t+1)$ or the $z$-transform
variable), that is, the discrete-time plant is
\beq
x(t+1) = A x(t) + B u(t),\;y(t) = C x(t),
\label{52eqn1d}
\eeq
and the discrete-time reference model system is
\beq
\dot{x}_m(t) = A_m x_m(t) + B_m u_m(t),\;y_m(t) = C_m x_m(t).
\label{52eqn1md}
\eeq

\medskip
With a unified symbol $D$ to denote either
$s$ or $z$, we can present a unified development of adaptive control
schemes for both the continuous-time and discrete-time cases.

\subsection{State Feedback Control Laws}
We first develop the nominal and adaptive state feedback output
tracking control laws using the reference model systems signals
$x_m(t)$ and $u_m(t)$.

\subsubsection{Nominal Control Law}
With the knowledge of the plant $(A, B, C)$, the nominal state
feedback control law is
\beq
u(t) = K_{1}^{*T} x(t) + K_{2}^* r_m(t),
\label{5ncm}
\eeq
where the parameter matrices $K_1^* \in R^{n \times M}$ and $K_2^*\in
R^{M \times M}$ satisfy
\beq
C (D I - A - B K_1^{* T})^{-1} B K_2^* = \xi_m^{-1}(D), K_2^{* -1} =
K_p,
\label{mepap3}
\eeq
whose existence is guaranteed \cite{wtl16}, and $r_m(t)$ is such that
\beq
y_m(t) = W_m(D)[r_m](t),\;W_m(D) = \xi_m^{-1}(D).
\eeq 

From (\ref{52eqn1m}), for $G_m(D) = C_m (D I - A_m)^{-1} B_m$, we have
\beq
y_m(t) = G_m(D)[u_m](t)
\eeq
so that $r_m(t) = \xi_m(D) G_m(D)[u_m](t)$ and $r_m(t)
= \xi_m(D)[y_m](t)$. Since $G_m(s)$ is unknown 
or the time-derivatives of $y_m(t)$ are not available, the signal
$r_m(t)$ is not available for implementing the control law (\ref{5ncm}),
and we need to parametrize its uncertainty for estimation.

\medskip
Recall the operator $D$ which is such that, for a scalar signal
$w(t)$, $D[w](t) = \dot{w}(t)$ in the continuous-time case or $D[w](t)
= w(t+1)$ in the discrete-time case. Then, under Assumption (A3.1):
$\lim_{s \rightarrow \infty} \xi_m(D) G_m(D)$ is finite, from the
reference system model (\ref{52eqn1m}), we can express
\beq
r_m(t) = \xi_m(D)[y_m](t) = A_1^T x_m(t) + A_2 u_m(t),
\label{5ximym}
\eeq
for some parameter matrices $A_1 \in R^{n \times M}$ and $A_2 \in R^{M
\times M}$ (similar to (\ref{5rm(t)u}) for the $M=1$ case), which
depend on the reference system parameters $(A_m, B_m, C_m)$ (the
matrix $A_2$ may not be nonsingular). Hence, we
can modify the nominal control law (\ref{5ncm}) as
\bea
u(t) \ts  = \ts K_{1}^{*T} x(t) + K_{2}^* r_m(t) \nn\\
\ts = \ts K_{1}^{*T} x(t) + K_{2}^* (A_1^T x_m(t) + A_2 u_m(t)) \nn\\
\ts = \ts K_{1}^{*T} x(t) + K_{21}^{*T} x_m(t) + K_{22}^* u_m(t), 
\label{5ncm1}
\eea
where $K_{21}^{*T} = K_{2}^* A_1^T,\;K_{22}^* = K_2^*A_2$.

Such a nominal control law ensures that $y(t) = W_m(D)[r_m](t) =
y_m(t)$ for the output $y_m(t)$ of a given reference model system: 
$D[x_m](t) = A_m x_m(t) + B_m u_m(t),\;y_m(t) = C_m x_m(t)$.

\subsubsection{Adaptive Control Law}
The adaptive controller structure is
chosen as
\beq
u(t) = K_{1}^{T} x(t) + K_{21}^{T} x_m(t) + K_{22} u_m(t), 
\label{5ncm1a}
\eeq
where $K_1$, $K_{21}$ and $K_{22}$ are the estimates of $K_1^*$,
$K_{21}^*$ and $K_{22}^*$.

\medskip
With the control law (\ref{5ncm1a}) applied to the unified form plant: 
\beq
D[x](t) = A x(t) + B u(t),\;y(t) = C x(t),
\label{5plantu}
\eeq
for the tracking error $e(t) = y(t) - y_m(t)$, we obtain
\beq
e(t) = W_m(D) K_p [(\Theta - \Theta^*)^T \omega](t),
\label{5e(t)su}
\eeq
where
\bea
\omega (t) \ts = \ts \left[x^{T}(t), x_m^T(t), u_m^T(t) \right]^{T} \in R^{2n+M}\\
\Theta (t) \ts = \ts \left[K_1^T(t), K_{21}^T(t),
K_{22}(t)\right]^T \in R^{(2n+M)\times M}\\
\Theta^{*} \ts = \ts \left[K_1^{*T}, K_{21}^{*T}, k_{22}^* \right]^T \in R^{(2n+M)\times M}.
\eea

\subsection{Output Feedback Control Laws}
The output feedback control laws uses either an observer-based
controller structure or a plant-model matching controller
structure. We now study their nominal form and adaptive version using
the signals $x_m(t)$ and $u_m(t)$ of the reference model system (\ref{52eqn1m}).

\subsubsection{Nominal Control Law}
The baseline nominal output feedback controller structure is
\begin{equation}
u(t) = \Theta_{1}^{*T}\omega_{1}(t) + \Theta_{2}^{*T}\omega_{2}(t) +
\Theta_{20}^* y(t) + \Theta_{3}^* r_m(t),
\label{542.18}
\end{equation}
where 
$\Theta_{1}^{*} = [\Theta_{11}^{*}, \ldots, \Theta_{1\,\nu-1}^{*}]^{T}$, 
$\Theta_{2}^{*} = [\Theta_{21}^{*}, \ldots, \Theta_{2\,\nu-1}^{*}]^{T}$, 
with $\Theta_{ij}^{*} \in R^{M \times M}$, $i = 1, 2$, $j = 1, 
\ldots, \nu-1$, with either $\nu = n-M$ or $\nu = \nu_p$ being the observability index of the
 plant (\ref{5plantu}), $\Theta_{20}^{*} \in R^{M
\times M}$ and $\Theta_{3} \in R^{M \times M}$, and
\beq
\omega_{1}(t) = F(D)[u](t),\;\omega_{2}(t) = F(D)[y](t),
\label{54omegas}
\eeq
\beq
F(D) = \frac{A(D)}{\Lambda(D)},\;A(D) = [I_M, D I_M, \ldots, D^{\nu - 
2} I_M]^{T},
\label{54F}
\eeq
for a chosen monic and stable polynomial $\Lambda(D)$ of degree
$\nu-1$. 

The nominal controller parameters $\Theta_1^*$, $\Theta_2^*$, $\Theta_{20}^{*}$
and $\Theta_{3}^{*}$ satisfy
\bea
\ts  \ts  \Theta_{1}^{*T} A(D)P(D) + (\Theta_{2}^{*T}A(D) + \Theta_{20}^* \Lambda(D))
 Z(D) \nn \\*[0.05in]
\ts  = \ts 
 \Lambda(D)(P(D) - \Theta_{3}^{*} \xi_{m}(D) Z(D)),
\label{54meq1}
\eea
for a pair of right matrix-fraction polynomial matrices $P(D)$ and $Z(D)$
of $G(D) = C(D\,I - A)^{-1} B$: $G(D) = Z(D) P^{-1}(D)$. With
(\ref{54meq1}), the closed-loop system transfer matrix is $W_m(s) =
\xi_m^{-1}(D)$.

\medskip
Using (\ref{5ximym}), we have
\bea
u(t) \ts = \ts \Theta_{1}^{*T}\omega_{1}(t) + \Theta_{2}^{*T}\omega_{2}(t) +
\Theta_{20}^* y(t) + \Theta_{3}^* (A_1^T x_m(t) + A_2 u_m(t)) \nn\\
\ts = \ts \Theta_{1}^{*T}\omega_{1}(t) + \Theta_{2}^{*T}\omega_{2}(t) +
\Theta_{20}^* y(t) + \Theta_{31}^{*T} x_m(t) + \Theta_{32}^* u_m(t)
\label{542.181}
\eea
where $\Theta_{31}^{*T} = \Theta_{3}^* A_1^T,\;\Theta_{32}^{*T}
= \Theta_{3}^* A_2$.

\begin{rem}
\rm
In the controller structure (\ref{542.18}), either $\nu = n-M$ (with
which the controller structure is observed-based, from using the
reduced-order estimate $\hat{x}(t)$ of $x(t)$ such that $\lim_{t \rightarrow \infty}
(\hat{x}(t) - x(t)) = 0$), or $\nu = \nu_p \leq n - M$ being the observability index of the
 plant (\ref{5plantu}) (with which the controller structure has the
 desired plant-model matching property (\ref{54meq1}) which makes the 
 closed-loop system transfer matrix equal to the reference model
 system $W_m(s) = \xi_m^{-1}(D)$). 
 \hspace*{\fill} $\Box$
\end{rem}

\subsubsection{Adaptive Control Law}
 We then choose the adaptive control law
\beq
u(t) =  \Theta_{1}^{T}\omega_{1}(t) + \Theta_{2}^{T}\omega_{2}(t) +
\Theta_{20} y(t) + \Theta_{31}^{T} x_m(t) + \Theta_{32} u_m(t)
\label{542.181a}
\eeq
with the estimates $\Theta_{1}$, $\Theta_{2}$, $\Theta_{20}$, $\Theta_{31}$ and
$\Theta_{32}$ of $\Theta_1^*$, $\Theta_2^*$, $\Theta_{20}^{*}$, $\Theta_{31}^{*}$ and
$\Theta_{32}^{*}$.

\medskip
With the control law (\ref{542.181a}) applied to the plant
(\ref{5plantu}), we can derive the dynamic equation for the tracking
error $e(t) = y(t) - y_m(t)$:
\beq
e(t) = W_m(D) K_p [(\Theta - \Theta^*)^T \omega](t),
\label{5e(t)ou}
\eeq
where
\bea
\omega (t) \ts = \ts \left[\omega_1^{T}(t), \omega_2^T(t), y^T(t), x_m^T(t), u_m^T(t) \right]^{T}\\
\Theta (t) \ts = \ts \left[\Theta_1^T(t), \Theta_{2}^T, \Theta_{20}, \Theta_{31}^T(t),
\Theta_{32}(t)\right]^T \\
\Theta^{*} \ts
= \ts \left[\Theta_1^{*T}, \Theta_2^{*T}, \Theta_{20}^*, \Theta_{31}^{*T}, \Theta_{32}^* \right]^T.
\eea
This error equation has the same form as that in (\ref{5e(t)su}) for
the state feedback control case, for which a unified adaptive law
design can be derived.

\subsection{Adaptive Laws}
The state feedback and output feedback control
designs lead to the tracking error equation (\ref{5e(t)su})
or (\ref{5e(t)ou}) which can be written in the form
\bea
\xi_m(D)[e](t) \ts = \ts K_p \tilde{\Theta}^T(t) \omega(t),\;\tilde{\Theta}(t)
= \Theta(t) - \Theta^* \nn\\
\ts = \ts K_p (u(t) - \Theta^{*T}(t) \omega(t)),\;u(t) = \Theta^T(t) \omega(t).
\label{5xime(t)}
\eea
Effectively dealing with the uncertainty of $K_p$ is a key step in
developing a stable adaptive scheme to update the controller parameter
matrix $\Theta(t)$, more sophisticated than the case of $M=1$ when
$K_p = k_p$ is a scalar. There are different adaptive schemes using
different information of $K_p$ \cite{t03}.

\subsubsection{A Basic Design}
We first present a basic adaptive scheme, using a gain matrix $S_p$
analogous to the sign of $k_p$ in the SISO case of $M=1$, specified in
the following assumption:

\begin{description}
\item[] {\bf Assumption (A3.3)}: A matrix $S_p \in R^{M \times M}$ is
known such that, for the continuous-time case, $K_{p} S_{p} = (K_{p}
S_{p})^{T} > 0$, and for the discrete-time
case, $0< K_{p} S_{p} = (K_{p} S_{p})^{T} < 2 I$ \cite{t03}.
\end{description}

For the first equation of (\ref{5xime(t)}), we choose a stable
and monic polynomial $f(D)$ of degree equal to the
maximum degree of $\xi_m(D)$, and operate both sides of
(\ref{5xime(t)}) by $h(D) = \frac{1}{f(D)}$, to obtain
\beq
\bar{e}(t) \stackrel{\triangle}{=} h(D) \xi_m(D)[e](t) = K_p
h(D)[\tilde{\Theta}^T \omega](t).
\label{54ebar}
\eeq

We then introduce the auxiliary signals
\beq
\zeta(t) = h(D)[\omega](t)
\label{54zeta}
\eeq
\beq
\xi(t) = \Theta^{T}(t) \zeta(t) - h(D)[\Theta^{T} \omega](t)
\label{54xi}
\eeq
and define the estimation error
\beq
\epsilon(t) = \bar{e}(t) + \Psi(t) \xi(t),
\label{54eps}
\eeq
where $\Psi(t)$ is the estimate of $\Psi^* = K_p$. It follows from 
(\ref{54ebar})-(\ref{54eps}) that
\begin{equation}
\epsilon(t) = K_{p}\tilde{\Theta}^{T}(t) \zeta(t) +
 \tilde{\Psi}(t) \xi(t),\;\tilde{\Psi}(t) = \Psi(t) - \Psi^*.
\end{equation}

\medskip
\medskip
{\bf Adaptive laws}. We choose the adaptive laws for $\Theta(t)$ and $\Psi(t)$:
\bea
\left.
\begin{array}{c}
\dot{\Theta}^{T}(t) \\
\Theta^T(t+1) - \Theta^T(t)  
\end{array}
\right\} 
\ts= \ts - \frac{S_{p}
\epsilon(t)\zeta^{T}(t)}{m^2(t)}\label{542.23}
\\*[0.05in]
\left.
\begin{array}{c}
\dot{\Psi}(t) \\
\Psi(t+1) - \Psi(t)
\end{array}
\right\} 
\ts = \ts - \frac{\Gamma \epsilon(t)\xi^{T}(t)}{m^2(t)},
\label{542.24}
\eea
where $S_p$ satisfies Assumption (A3.3), $\Gamma = \Gamma^T >
0$ for the continuous-time case and $0 < \Gamma = \Gamma^T < 2 I$ for
the discrete-time case, and $m(t) = \sqrt{1 + \zeta^T(t) \zeta(t) + \xi^T(t)
\xi(t)}$.

\medskip
\medskip
{\bf Stability analysis}. Consider the the positive definite function
\begin{equation}
V = \tr[\tilde{\Theta} \Gamma_{p}
 \tilde{\Theta}^{T}] + \tr[\tilde{\Psi}^{T} \Gamma^{-1}
 \tilde{\Psi}],\;\Gamma_{p} = K_{p}^{T} S_{p}^{-1} = \Gamma_p^T > 0,
\label{5Vu}
\end{equation}
we derive, for the continuous-time case, its time-derivative as
\bea
\dot{V} \ts = \ts 2 \tr[\tilde{\Theta} \Gamma_{p}
 \dot{\Theta}^{T}] + 2 \tr[\tilde{\Psi}^{T} \Gamma^{-1}
 \dot{\Psi}] \nn\\
\ts = \ts - 2\tr[\tilde{\Theta} \Gamma_{p}
  \frac{S_{p} \epsilon(t)\zeta^{T}(t)}{m^2(t)}] - 2 \tr[\tilde{\Psi}^{T} \Gamma^{-1}
 \frac{\Gamma \epsilon(t)\xi^{T}(t)}{m^2(t)}] \nn\\
\ts = \ts - 2\tr[\zeta^T(t) \tilde{\Theta} \Gamma_{p} S_p 
  \frac{\epsilon(t)}{m^2(t)}] -
  2 \tr[\xi^T(t) \tilde{\Psi}^{T} \frac{\epsilon(t)}{m^2(t)}] \nn\\
\ts = \ts -2 (\zeta^T(t) \tilde{\Theta} K_p^T
  + \xi^T(t) \tilde{\Psi}^{T}) \frac{\epsilon(t)}{m^2(t)} \nn\\
\ts = \ts - \frac{2 \epsilon^{T}(t) \epsilon(t)}{m^2(t)} \leq 0,
\eea
or, for the discrete-time case, its time-increment as 
\bea
\ts \ts V(\tilde{\Theta}(t+1), \tilde{\Psi}(t+1))-
V(\tilde{\Theta}(t), \tilde{\Psi}(t)) \nn\\
\ts = \ts 
\tr[\tilde{\Theta}(t+1) \Gamma_{p} \tilde{\Theta}^{T}(t+1)]
+ \tr[\tilde{\Psi}^{T}(t+1) \Gamma^{-1} \tilde{\Psi}(t+1)] \nn\\
\ts \ts 
- \tr[\tilde{\Theta}(t) \Gamma_{p} \tilde{\Theta}^{T}(t)]
- \tr[\tilde{\Psi}^{T}(t) \Gamma^{-1} \tilde{\Psi}(t)] \nn\\
\ts = \ts 
\tr[(\tilde{\Theta}^T(t)
  - \frac{S_{p} \epsilon(t)\zeta^{T}(t)}{m^2(t)})^T \Gamma_p (\tilde{\Theta}^T(t)
  - \frac{S_{p} \epsilon(t)\zeta^{T}(t)}{m^2(t)})] 
- \tr[\tilde{\Theta}(t) \Gamma_{p} \tilde{\Theta}^{T}(t)] \nn\\
\ts \ts 
+ \tr[(\tilde{\Psi}(t)
- \frac{\Gamma \epsilon(t)\xi^{T}(t)}{m^2(t)})^T \Gamma^{-1} 
(\tilde{\Psi}(t) - \frac{\Gamma \epsilon(t)\xi^{T}(t)}{m^2(t)})] 
- \tr[\tilde{\Psi}^{T}(t) \Gamma^{-1} \tilde{\Psi}(t)] \nn\\
\ts = \ts - 2\tr[\frac{\zeta(t) \epsilon^T(t)
S_p^T \Gamma_p \tilde{\Theta}^T(t)}{m^2(t)}] + \tr[\frac{\zeta(t) \epsilon^T(t)
S_p^T \Gamma_p S_p \epsilon(t) \zeta^T(t)}{m^4(t)}] \nn\\
\ts \ts - 2\tr[\frac{\xi(t) \epsilon^T(t)
\Gamma \Gamma^{-1} \tilde{\Psi}(t)}{m^2(t)}] + \tr[\frac{\xi(t) \epsilon^T(t)
\Gamma \Gamma^{-1} \Gamma \epsilon(t) \xi^T(t)}{m^4(t)}] \nn\\
\ts=\ts - \frac{2 \epsilon^T(t) \epsilon(t)}{m^2(t)} +
\frac{\zeta^T(t) \zeta(t)}{m^2(t)} \frac{\epsilon^T(t) K_p S_p \epsilon(t)}{m^2(t)}
+ \frac{\xi^T(t) \xi(t)}{m^2(t)} \frac{\epsilon^T(t) \Gamma \epsilon(t)}{m^2(t)} \nn\\
\ts \leq \ts - \frac{\alpha_1 \epsilon^T(t) \epsilon(t)}{m^2(t)},
\eea
for $0< \alpha_1 \leq 2
- \max\{\lambda_{max}[\Gamma_p], \lambda_{max}[\Gamma]\}$. 

\medskip
Based on this result, we can derive the following properties:

\begin{lem}
\label{542.23ps}
The adaptive laws (\ref{542.23})-(\ref{542.24}) ensure:

\medskip
(i) $\Theta(t) \in L^\infty$, $\Psi(t) \in L^\infty$, and 
and $\frac{\epsilon(t)}{m(t)} \in L^2 \cap L^{\infty}$; 

\medskip
(ii) $\dot{\Theta}(t) \in  L^2 \cap L^\infty$ and $\dot{\Psi}(t) \in  L^2 \cap
L^\infty$ (the continuous-time case); and 

\medskip
(iii) $\Theta(t+1)-\Theta(t) \in  L^2$ and $\Psi(t+1) - \Psi(t) \in
L^2$ (the discrete-time case).
\end{lem}


\subsubsection{$K_p$ Decomposition Based Designs}
The knowledge of the plant high frequency gain matrix $K_p$, needed
for adaptive control design, can be reduced by using the LDS, LDU and
SDU decompositions of $K_p$ \cite{t03}. 

\medskip
The LDS based design can be derived, by reparametrizing the first
equation of (\ref{5xime(t)}) as
\beq
\xi_m(D)[e](t) + \Theta_0^* \xi_{m}(D)[e](t) = D_s S \tilde{\Theta}^T(t)\omega(t),
\label{52e(t)2}
\eeq
where $K_p = L_s D_s S$ with $L_s$ being a unity lower triangular
matrix ($D_s$ being a diagonal matrix and $S$ being a symmetric and
positive definite matrix), and $\Theta_0^* = L_s^{-1} - I_M$ has the
special structure:
\begin{eqnarray}
\Theta_0^* = \left[ \begin{array}{ccccc}
0 & 0 & 0 &  \cdots & 0 \\
\theta_{21}^*      & 0  & 0 & \cdots & 0\\
\theta_{31}^*      & \theta_{32}^*  & 0 & \cdots & 0\\
           & \cdots           & \cdots    & \cdots &   \\
\theta_{M-1\,1}^*  & \cdots & \theta_{M-1\,M-2}^* & 0 & 0 \\
\theta_{M\,1}^*  & \cdots & \theta_{M\,M-2}^* & \theta_{M\,M-1}^* & 0 
\end{array}
\right].
\label{24Theta_0*1}
\end{eqnarray}

\medskip
The LDU decomposition of $K_p$: $K_p = L D_p U$, with $L$ being 
a unity lower triangular matrix and $U$ being a unity upper triangular
matrix (and $D_p$ being a diagonal matrix), can also be used for an
adaptive design. In this case, the second equation of (\ref{5xime(t)})
is reparametrized as
\bea
\xi_m(D)[e](t) + \Theta_0^* \xi_{m}(D)[e](t) \ts = \ts D_p U (u(t)
- \Theta^{*T}\omega(t)) \nn\\
\ts = \ts D_p (u(t) - \Phi_0^* u(t) - U \Theta^{*T} \omega(t)),
\label{52e(t)21}
\eea
where $\Phi_0^* = I - U = \left[ \begin{array}{ccccc}
0 & \phi_{12}^{*} & \phi_{13}^{*} &  \cdots & \phi_{1M}^{*} \\
0      & 0  & \phi_{23}^{*} & \cdots & \phi_{2M}^{*} \\
\vdots      & \vdots           & \vdots    & \vdots & \vdots  \\
0      & 0 & \cdots & 0 & \phi_{M-1\,M}^{*} \\
0          & \cdots           & \cdots           & 0      & 0
\end{array}
\right]$ whose elements $\phi_{ij}^*$ can be estimated.

\medskip
The parameters $\Theta_0^*$, $D_s$ and $\Theta^*$ (in
$\tilde{\Theta}(t) = \Theta(t) - \Theta^*$) in (\ref{52e(t)2})
 can be estimated, using the sign information of the diagonal elements
 of $D_s$. The parameters $\Theta_0^*$, $\Phi_0^*$, $U \Theta^{*T}$
 and $D_p = \diag\{d_1^*, d_2^*, \ldots, d_M^*\}$ in (\ref{52e(t)21})
 can be estimated, using the sign information of
 $d_i^*$. Such sign information depends on the leading principal
 minors of $K_p$, which are assumed to be all non-zero. For the
 discrete-time case design, some bound information $d_i^0 \geq
 |d_i^*|$ is needed for the adaptive law. For details of such $K_p$
 decomposition based adaptive control designs, see \cite[pages 448-452]{t03}.

\subsection{Further Extensions}
We now present some additional designs of the adaptive control schemes
for the reference model system (\ref{52eqn1m}) with uncertain $(A_m,
B_m, C_m)$, and their performance analysis and applications.

\subsubsection{Design with $x_m(t)$ Unavailable}
When the state vector $x_m(t)$ of the reference system (\ref{52eqn1m}):
\beq
D[x_m](t) = A_m x_m(t) + B_m u_m(t),\;y_m(t) = C_m x_m(t),
\label{5rsu}
\eeq
is not available, a nominal reduced-order state observer \cite[page 272]{r96}
can be used, based on the knowledge of $y_m(t) \in R^M$ and
$u_m(t) \in R^M$, to generate an estimate $\hat{x}_{m}(t)$ of $x_m(t)$ such that 
$\lim_{t \rightarrow \infty} (\hat{x}_{m}(t) - x_m(t)) = 0$
exponentially. With such an $\hat{x}_m(t)$, the term 
$A_1^T x_m(t)$ in (\ref{5ximym}) can be replaced with 
$A_1^T \hat{x}_m(t)$ which can be further expressed as
\beq
A_1^T \hat{x}_m(t) = B_{1}^{T}\omega_{u_m}(t) + B_{2}^{T}\omega_{y_m}(t) +
B_{20} y_m(t),
\eeq
for some constant matrices $B_1$, $B_2$, $B_{20}$, and
$
\omega_{u_m}(t) = F_m(D)[u_m](t)$, $\omega_{y_m}(t) = F_m(D)[y_m](t),
$
with $
F_m(D) = \frac{A_m(D)}{\Lambda_e(D)},\;A_m(D) = [I_M, D I_M, \ldots,
D^{n-M-1} I_M]^{T},
$
for a chosen monic and stable polynomial $\Lambda_e(D)$ of degree
$n-M-1$. 

We then can express $r_m(t)$ in (\ref{5ximym}) as
\beq
r_m(t) = B_1^T \omega_{u_m}(t) + B_{2}^T \omega_{y_m}(t) +
B_{20} y_m(t) + A_2 u_m(t),
\label{5r02u1}
\eeq
which can be combined with the control law (\ref{5ncm}) or
(\ref{542.18}) to derive a control law in terms of the reference model 
system signals $u_m(t)$ and $y_m(t)$.

\subsubsection{Partial-State Feedback Control Designs}
In \cite{st21b}, a partial-state feedback  adaptive control scheme is
developed for MIMO systems to achieve output tracking, using an
available partial-state vector $y_0(t) = C_0 x(t) \in R^{n_0}$ with
$(A, C_0)$ observable and $\rank[C_0] = n_0$, to build an
observer-based control law. In \cite{t24b}, a partial-state
feedback adaptive control scheme with uncertain reference model system
parameters is developed for SISO systems. 

\medskip
For a MIMO plant, the baseline nominal partial-state feedback controller structure
\cite{st21b} is
\begin{align} \label{11NominalControl}
u(t) & = \Theta_1^{*T}\omega_1(t) + \Theta_2^{*T}\omega_2(t)
+ \Theta_{20}^{*}y_0(t) + \Theta_3^*r_m(t), 
\end{align}
where $\Theta_{1}^* \in
R^{M(n-n_0) \times M}$, $\Theta_2^* \in R^{n_0(n-n_0) \times M}$,
$\Theta_{20}^* \in R^{M \times n_0}$ and $\Theta_3^* \in R^{M \times
M}$, and 
\beq
\omega_1(t) = \frac{A_1(D)}{\Lambda(D)}[u](t), \; \omega_2(t)
= \frac{A_2(D)}{\Lambda(D)}[y_0](t)
\eeq
with $A_1(D) = [I_{M},
D I_{M},\ldots, D^{n-n_0-1}I_{M}]^T$ and $A_2(D)=[I_{n_0}, D I_{n_0},\ldots,
  D^{n-{n_0}-1}I_{n_0}]^T$, and $\Lambda(D)$ being a monic stable
polynomial of degree $n - n_0$.

\medskip
In our control problem, $r_m(t) = \xi_m(D)[y_m](t)$ is not
available and we reparametrize it. With the parametrization scheme
(\ref{5ximym}): $r_m(t) = \xi_m(D)[y_m](t) = A_1^T x_m(t) + A_2
u_m(t)$, we express
\beq
\Theta_3^*r_m(t) = \Theta_3^*(A_1^T x_m(t) + A_2 u_m(t))
\stackrel{\triangle}{=} \Theta_{31}^{*T} x_m(t) + \Theta_{32}^{*}
u_m(t),
\label{Theta3rmxm}
\eeq
to form the nominal control law (which uses $x_m(t)$ and $u_m(t)$):
\beq
u(t) = \Theta_1^{*T}\omega_1(t) + \Theta_2^{*T}\omega_2(t)
+ \Theta_{20}^{*}y_0(t) + \Theta_{31}^{*T} x_m(t) + \Theta_{32}^{*}
u_m(t),
\eeq
whose adaptive version is
\beq
u(t) = \Theta_1^{T}\omega_1(t) + \Theta_2^{T}\omega_2(t)
+ \Theta_{20} y_0(t) + \Theta_{31}^{T} x_m(t) + \Theta_{32} u_m(t),
\eeq
with the estimates $\Theta_1$, $\Theta_2$, $\Theta_{20}$,
$\Theta_{31}$ and $\Theta_{32}$ of $\Theta_1^*$, $\Theta_2^*$, $\Theta_{20}^*$,
$\Theta_{31}^*$ and $\Theta_{32}^*$.

\medskip
With the parametrization scheme (\ref{5r02u1}) (using the
observer-based estimate $\hat{x}_m(t)$ of $x_m(t)$): 
$r_m(t) = B_1^T \omega_{u_m}(t) + B_{2}^T \omega_{y_m}(t) +
B_{20} y_m(t) + A_2 u_m(t)$, we express 
\bea
\Theta_3^*r_m(t) \ts = \ts \Theta_3^* (B_1^T \omega_{u_m}(t) + B_{2}^T \omega_{y_m}(t) +
B_{20} y_m(t) + A_2 u_m(t)) \nn\\
\ts \stackrel{\triangle}{=} \ts \Theta_{31}^{*T} \omega_{u_m}(t) + \Theta_{32}^{*T} \omega_{y_m}(t) +
\Theta_{33}^{*} y_m(t) + \Theta_{34}^{*} u_m(t),
\label{Theta3rmym}
\eea
to form the nominal control law (which uses $y_m(t)$ and $u_m(t)$):
\beq
u(t) = \Theta_1^{*T}\omega_1(t) + \Theta_2^{*T}\omega_2(t)
+ \Theta_{20}^{*}y_0(t) + \Theta_{31}^{*T} \omega_{u_m}(t) + \Theta_{32}^{*T} \omega_{y_m}(t) +
\Theta_{33}^{*} y_m(t) + \Theta_{34}^{*} u_m(t),
\eeq
whose adaptive version is 
\beq
u(t) = \Theta_1^{T}\omega_1(t) + \Theta_2^{T}\omega_2(t)
+ \Theta_{20}y_0(t) + \Theta_{31}^{T} \omega_{u_m}(t) + \Theta_{32}^{T} \omega_{y_m}(t) +
\Theta_{33} y_m(t) + \Theta_{34} u_m(t),
\eeq
with the estimates $\Theta_1$, $\Theta_2$, $\Theta_{20}$,
$\Theta_{31}$, $\Theta_{32}$, $\Theta_{33}$ and $\Theta_{34}$ of
$\Theta_1^*$, $\Theta_2^*$, $\Theta_{20}^*$, $\Theta_{31}^*$,
$\Theta_{32}^*$, $\Theta_{33}^*$ and $\Theta_{34}^*$.

\medskip
With these adaptive control laws, the tracking error equations can be
derived and the estimation error signals can be defined to design the
adaptive laws to update the controller parameters.

\subsubsection{Design for Continuous-Time Relative-Degree-One Systems}
A continuous-time relative-degree-one system is 
\beq
\dot{x}(t) = A x(t) + B u(t),\;y(t) = C x(t),
\label{5plantc}
\eeq
with the modified interactor matrix $\xi_m(s) = s I + P_0$ for 
$P_0 = \diag\{a_1, a_2, \ldots, a_M\} > 0$.

\medskip
In this case, the state feedback tracking error equation
(\ref{5e(t)su}) and the output feedback tracking error equation
(\ref{5e(t)ou}) have the form
\beq
\dot{e}(t) = A_0 e(t) + K_p (\Theta - \Theta^*)^T(t) \omega(t),
\label{5e(t)sc1}
\eeq
where $A_0 = - P_0 = \diag\{-a_1, -a_2, \ldots, -a_M\}$ with $a_i
> 0$, $i = 1, 2, \ldots, M$, that is, $A_0$ is a stable matrix. In
this case, a Lyapunov design of the adaptive
law for $\Theta(t)$ can be derived, and the closed-loop system
stability and tracking performance can be analyzed in a
straightforward way.

\medskip
\medskip
{\bf Adaptive law}. We choose the adaptive law
\beq
\dot{\Theta}^{T}(t) = - S^T P e(t) \omega^T(t),
\label{5Thetadot1}
\eeq
where $P = P^{T} > 0$ satisfying $P A_{0} + A_{0}^{T} P = - Q$ for $Q
= Q^{T} > 0$ chosen, and $S_p \in R^{M \times M}$ is such that $M_{s}
= K_{p}^{-1} S = M_s^T > 0$ (the knowledge of $S$ needs to be
assumed for such an adaptive law, which may be relaxed using an LDU
decomposition of $K_p$ \cite[page 375]{t03}).

\medskip
\medskip
{\bf Stability analysis}. Consider the positive definite function
\beq
V = e^{T} P e + \tr[\tilde{\Theta} M_{s}^{-1} \tilde{\Theta}^{T}], \;
\tilde{\Theta} = \Theta - \Theta^*, 
\label{5Vc1}
\eeq
and derived its time-derivative as
\bea
\dot{V} \ts = \ts 2 e^T(t) P (A_0 e(t) + K_p \tilde{\Theta}^T 
\omega(t)) - 2 \tr[\tilde{\Theta} M_{s}^{-1} S^T P
  e(t) \omega^T(t)] \nn\\
\ts = \ts - e^T(t) Q e(t) + 2 e^T(t) K_p \tilde{\Theta}^T(t)
 \omega(t)) - 2 \omega^T(t) \tilde{\Theta} M_{s}^{-1}
  S^T P e(t) \nn\\
\ts = \ts - e^T(t) Q e(t),
\eea
Hence, $e(t) \in L^2$ for $e(t) = y(t) - y_m(t)$, and 
$e(t)$ and $\tilde{\Theta}(t) = \Theta(t) - \Theta^*$ are bounded, and
so are $y(t)$ and $\Theta(t)$. It can be further shown that $u(t)$ is
bounded, and $\lim_{t \rightarrow \infty} e(t) = 0$.

\subsubsection{Asymptotic Tracking Performance}
The control objective is to ensure closed-loop signal boundedness and
asymptotic output tracking: $\lim_{t \rightarrow \infty} (y(t) -
y_m(t)) = 0$, without the knowledge of the equivalent reference system
input $r_m(t) = \xi_m(D) G_m(D)[u_m](t)$ as $G_m(D)$ is unknown for
the reference model system (\ref{52eqn1m}) or (\ref{5rsu}), 
which can be achieved by the control laws (\ref{5ncm1a})
and (\ref{542.181a}). 

The control scheme with (\ref{5ncm1a}) has the
terms $K_{21}^{T} x_m(t) + K_{22} u_m(t)$ replacing $K_2 r_m(t)$ in 
a standard adaptive state feedback control
scheme \cite{t03}. Similarly, the scheme with (\ref{542.181a})
has the terms $\Theta_{31}^{T} x_m(t) + \Theta_{32} u_m(t)$ replacing
$\Theta_3 r_m(t)$ in an adaptive output feedback control
scheme \cite{t03}. Since these terms are all bounded, the closed-loop
system signal boundedness is ensured. Then, the $L^2$ properties in
Lemma \ref{542.23ps}, ensure the $L^2$ property of the tracking error
$e(t) = y(t) - y_m(t)$ and the asymptotic convergence property of $e(t)$.

\subsubsection{Applications}
The developed new adaptive control technique ensures that the output
$y(t)$ of the plant: $D[x](t) = A x(t) + B u(t),\;y(t) = C x(t)$, 
tracks the output $y_m(t)$ of the reference system: $D[x_{m}](t) =
A_{m} x_{m}(t) + B_{m} u_m(t),\;y_m(t) = C_m x_m(t)$ (with known
$y_m(t)$ and $u_m(t)$, and possibly $x_m(t)$) whose parameters
($A_m, B_m, C_m$) are unknown, while in the traditional model reference
control framework, the reference model system $y_m(s) =
W_m(D)[r_m](t)$ is assumed to be completely known.

An immediate application is adaptive control of a follower system with
unknown parameters to follow a leader system also with unknown parameters. 

\medskip
The adaptive control technique, developed for the leader-follower
control problem (with the reference model system with unknown
parameters being the leader), can be generalized for adaptive control of multi-agent
systems whose leader system parameters are unknown, in addition to the
uncertainties of the follower agent systems. It is meaningful to
consider the parameter uncertainties of the leader agent for an
adaptive multi-agent control problem.

\setcounter{equation}{0}
\section{Adaptive Feedback Linearization Control}
Consider a nonlinear plant of the form
\bea
\dot{x}(t) \ts = \ts \sum_{i=1}^{l}\theta_i^*
f_i(x)+ \sum_{i=1}^M g_i(x)u_i(t) =  F(x) \theta^* + G(x) u\nn\\
y \ts = \ts h(x) = [h_1(x), h_2(x), \ldots, h_M(x)]^T \in R^M,
\label{6fsystem}
\eea
with $\theta^* = [\theta_1^*, \theta_2^*,
\ldots, \theta_l^*]^T \in R^l$ unknown, and $F(x) = [f_1(x), f_2(x), \ldots,
f_l(x)] \in R^{n \times l}$ and $G(x) = [g_1(x), g_2(x), \ldots,
g_M(x)] \in R^{n \times M}$ known, and assume that the plant has a vector
relative degree $\diag \{\rho_1, \rho_2, \ldots, \rho_M\}$ and stable
zero dynamics \cite{i95}.

The control objective is to design a feedback control signal $u(t) \in
R^M$ for (\ref{6fsystem}), to ensure the control system signal
boundedness and asymptotic tracking of a bounded reference output signal
$y_m(t) \in R^M$ by the plant output $y(t)$.

In a traditional adaptive feedback linearization control system
\cite{sb89}, \cite{si89}, \cite{wts20}, \cite{wtyz16}, the reference
output signal $y_m(t)$ is a given signal whose certain orders of
time-derivatives are assumed to be bounded and known. In out study,
$y_m(t)$ is the output of a reference model system (a leader system)
whose parameters are unknown.

\subsection{Feedback Linearization Parametrization}
We first give a review of the standard feedback linearization
parametrization procedure.

\medskip
Letting $\xi(s) = \diag\{s^{\rho_1}, s^{\rho_2}, \ldots, s^{\rho_M}\}$,
and following the standard feedback linearization
procedure \cite{sb89}, we can derive the plant expression
\beq
\xi(s)[y](t) = b(x) + A(x) u(t),
\label{66aplantfl}
\eeq
for some parametrizable $b(x) \in R^n$ and $A(x) \in
R^{M \times M}$ being nonsingular. 

\medskip
With $\hat{A}(x)$ and $\hat{b}(x)$ being the estimates (to be
parametrized) of $A(x)$ and $b(x)$, we choose the adaptive linearizing control law
\beq
u = (\hat{A}(x))^{-1} (v - \hat{b}(x)).
\label{6lcontrol}
\eeq
Then, we have $\hat{A}(x) u = v - \hat{b}(x)$, and 
\bea
\ts \ts b(x) + A(x) u \nn\\
\ts = \ts A(x) u - \hat{A}(x) u + v + b(x) - \hat{b}(x),
\label{6c.11}
\eea
which leads the plant expression (\ref{66aplantfl}) to
\beq
\xi(s)[y](t) = A(x) u - \hat{A}(x) u + b(x) - \hat{b}(x) + v.
\label{6fl}
\eeq

For $\xi_m(s) = \diag\{d_1(s), d_2(s), \ldots, d_M(s)\}$ with 
$d_i(s)=s^{\rho_i}+\alpha_{i1}s^{\rho_i-1}+\cdots+\alpha_{i\rho_i-1} s
+\alpha_{i\rho_i}$ stable, $i=1, 2, \cdots, M$, and $v = [v_1,
v_2, \ldots, v_M]^T$, we choose 
\bea
v_i \ts =\ts y_{mi}^{(\rho_i)}+
\alpha_{i1}(y_{mi}^{(\rho_i-1)}-\widehat{L_f^{\rho_i-1}h_i(x)})+\cdots
+ 
\alpha_{i\rho_i-1}(y_{mi}^{(1)}-\widehat{L_f^{1}h_i(x)})+ \alpha_{i\rho_i}(y_{mi}-y_{i}),
\label{6c.46}
\eea
as the adaptive control law, 
where $\widehat{L_f^{\rho_i-1}h_i(x)}={L_{\hat{f}}^{\rho_i-1}h_i(x)},
\ldots, \widehat{L_f^{1}h_i(x)}={L_{\hat{f}}^{1}h_i(x)}$ 
are parametrized by the estimates of $\theta_i^*$ in the plant
(\ref{6fsystem}), and $y_{mi}(t)$, $i=1,2,\ldots, M$, are the components
of the reference output vector signal $y_m(t) = [y_{m1}(t), y_{m2}(t),
\ldots, y_{mM}(t)]^T \in R^M$, which and whose 
time-derivatives of certain orders as shown in (\ref{6c.46}) are
assumed to be bounded and known.

\medskip
Recall the Lie Derivatives defined as $L^k_f h_i=L_f(L^{k-1}_f
h_i)=\frac{\partial{L^{k-1}_f h_i}}{\partial{x}}f,\; L_f^0h_i=h_i,\;
L_{g_j}L^k_f h_i=\frac{\partial{L^{k}_f h_i}}{\partial{x}}g_j$
\cite{i95}. 
Using the control law (\ref{6c.46}) in (\ref{6c.11}), for $e = y(t) -
y_m(t) = [e_1,
e_2, \ldots, e_M]^T$ with $e_i = y_i - y_{mi}$, we obtain the error equation
\bea
\ts\ts \left[\begin{array}{c}
e_{1}^{(\rho_1)}+\alpha_{11}e_{1}^{(\rho_1-1)}+\cdots+\alpha_{1\rho_1}e_{1}\\
e_{2}^{(\rho_2)}+\alpha_{21}e_{2}^{(\rho_2-1)}+\cdots+\alpha_{2\rho_2}e_{2}\\
\vdots \\
e_{M}^{(\rho_M)}+\alpha_{M1}e_{M}^{(\rho_M-1)}+\cdots+\alpha_{M\rho_M}e_{M}
\end{array}\right] =\tilde{b}(x)+\tilde{A}(x) u(t) \nn\\
\ts + \ts 
\left[\begin{array}{c}
\alpha_{11}(y_{1}^{(\rho_1-1)}-\widehat{L_f^{\rho_1-1}h_1(x)})+\cdots+\alpha_{1\rho_1-1}(\dot{y}_{1}-\widehat{L_f h_{1}(x)})\\
\alpha_{21}(y_{2}^{(\rho_2-1)}-\widehat{L_f^{\rho_2-1}h_2(x)})+\cdots+\alpha_{2\rho_2-1}(\dot{y}_{2}-\widehat{L_f h_{2}(x)})\\
\vdots \\
\alpha_{M1}(y_{M}^{(\rho_M-1)}-\widehat{L_f^{\rho_M-1}h_M(x)})+\cdots+\alpha_{M\rho_M-1}(\dot{y}_{M}-\widehat{L_f h_{M}(x)})
\end{array}\right]\!,\;\;\;\;\;\;\;\;\;
\label{6c.24}
\eea
where $y_i^{(\rho_i-1)} = L_{f}^{\rho_i-1}h_i(x),
\ldots, y_i^{(1)} = L_{f}^{1}h_i(x)$ are parametrized by the parameters in $\theta^*$,
$\tilde{b}(x)=b(x)-\hat{b}(x)$ and  
$\tilde{A}(x)=A(x)-\hat{A}(x)$.

\medskip
Parametrizing $b(x)=\Theta_1^{*T}\omega_1$ and 
$\hat{b}(x)=\Theta_1^T\omega_1$ 
for some known vector $\omega_1$ and
unknown parameter matrix $\Theta_1^*$ and its estimate $\Theta_1$, 
$A(x) u(t)= \Theta_2^{*T}\omega_2$ and
 $\hat{A}(x) u(t)= \Theta_2^T\omega_2$ for some known vector $\omega_2$ and
unknown parameter matrix $\Theta_2^*$ and its estimate $\Theta_2$, we
express
\beq
\tilde{b}(x)=\tilde{\Theta}_1^T\omega_1,\;\tilde{\Theta}_1 =
\Theta_1^* - \Theta_1
\eeq
\beq
\tilde{A}(x) u(t)=\tilde{\Theta}_2^T\omega_2, \;\tilde{\Theta}_2 =
\Theta_2^* - \Theta_2. 
\eeq
Similarly, we can express the last term in (\ref{6c.24}) as
$\tilde{\Theta}_3^T\omega_3$ for some known vector $\omega_3$ and
$\tilde{\Theta}_3 = \Theta_3^* - \Theta_3$ with some unknown parameter
matrix $\Theta_3^*$ and its estimate $\Theta_3$. Then, denoting 
$\Theta^{*}=[\Theta_1^{*T}, \Theta_2^{*T}, \Theta_3^{*T}]^T$, 
$\Theta=[\Theta_1^{T}, \Theta_2^{T}, \Theta_3^{T}]^T$,
$\tilde{\Theta}=\Theta^*-\Theta$ and $\omega=[\omega_1^T, \omega_2^T,
\omega_3^T]^T$, we rewrite (\ref{6c.24}) as
\bea
\ts \ts \left[\begin{array}{c}
e_{1}^{(\rho_1)}+\alpha_{11}e_{1}^{(\rho_1-1)}+\cdots+\alpha_{1\rho_1}e_{1}\\
e_{2}^{(\rho_2)}+\alpha_{21}e_{2}^{(\rho_2-1)}+\cdots+\alpha_{2\rho_2}e_{2}\\
\vdots \\
e_{M}^{(\rho_M)}+\alpha_{M1}e_{M}^{(\rho_M-1)}+\cdots+\alpha_{M\rho_M}e_{M}
\end{array}\right] \nn\\
\ts=\ts   \tilde{\Theta}_1^T \omega_1+ \tilde{\Theta}_2^T \omega_2 +
\tilde{\Theta}_3^T \omega_3 = \tilde{\Theta}^T\omega.
\label{6c.33}
\eea

Thus, we have expressed the adaptive control error system as
\beq
\xi_m(s)[e](t) = (\Theta^* - \Theta(t))^T \omega(t).
\label{6etheta-thetas}
\eeq

Based on such an error system, adaptive laws can be developed to
update the parameter estimate $\Theta(t)$ to implement the adaptive
control scheme with the control laws (\ref{6lcontrol}) and
(\ref{6c.46}), under the stable zero dynamics condition of the plant
(\ref{6fsystem}), using the knowledge of the certain
orders of derivatives of $y_m(t)$, that is, $\xi_m(s)[y_m](t)$, which
is however not available in our adaptive control problem.

\subsection{Leader-Follower Tracking Error Equation}
Now we develop the standard adaptive feedback linearization control
design, with an expanded parametrization, for the adaptive
leader-follower tracking problem: the output $y(t)$ of
the plant (\ref{6fsystem}) (the follower) with unknown parameters is
expected to track the output $y_m(t)$ of a leader system (a reference
model system) with unknown parameters.

\medskip
{\bf Reference system}. Assume that $y_m(t)$ is the output of a
reference model system:
\bea
\dot{x}_m \ts = \ts f_m(x_m) + g_m(x) u_m\nn\\
y_m \ts = \ts h_m(x_m),
\label{6rsf}
\eea
with the vector relative degree no less than $\diag \{\rho_1, \rho_2,
\ldots, \rho_M\}$ of the follower system (\ref{6fsystem}), that
is, for $\xi(s) = \diag\{s^{\rho_1}, s^{\rho_2}, \ldots,
s^{\rho_M}\}$, it holds that
\beq
\xi(s)[y_m](t) = b_{m0}(x_m) + A_m(x_m) u_m(t),
\label{66aplantflm}
\eeq
for some $b_{m0}(x) \in R^n$ and $A_m(x) \in R^{M \times M}$ (which
may be singular), and
\beq
\xi_m(s)[y_m](t) = b_{m}(x_m) + A_m(x_m) u_m(t),
\label{66aplantflm1}
\eeq
for some $b_m(x) \in R^n$, where $\xi_m(s) = \diag\{d_1(s), d_2(s),
\ldots, d_M(s)\}$ with $d_i(s)=s^{\rho_i}+\alpha_{i1}s^{\rho_i-1}+\cdots+\alpha_{i\rho_i-1} s
+\alpha_{i\rho_i}$ stable, $i=1, 2, \cdots, M$.

\medskip
To design the adaptive control schemes, we make the following assumption:

\begin{description}
\item[] {\bf Assumption (A4.1)}: the reference system (leader system)
signals $u_m(t)$, $x_m(t)$ and $y_m(t)$ are known and bounded but the
dynamic functions $f_m(\cdot)$, $g_m(\cdot)$ and $h_m(\cdot)$ are
unknown.
\end{description}

Our objective is to design an adaptive control law $u(t)$ for the
follower system (\ref{6fsystem}), to ensure that all system signals are
bounded, and $\lim_{t \rightarrow \infty} (y(t) - y_m(t)) = 0$,
without the information of $\xi_m(s)[y_m](t)$ (except that $y_m(t)$ is 
available and $\xi_m(s)$ is chosen).

\medskip
The control law (\ref{6c.46}) can be written as
\beq
v(t) = \xi_m(s)[y_m](t) - \hat{v}_y(x),
\label{6c.461}
\eeq
where
\beq
\hat{v}_y(x) = \left[\begin{array}{c}
\alpha_{11} \widehat{L_f^{\rho_1-1}h_1(x)}+\cdots + 
\alpha_{1\rho_1-1} \widehat{L_f^{1}h_1(x)}+ \alpha_{1\rho_1} y_{1} \\
\alpha_{21} \widehat{L_f^{\rho_2-1}h_2(x)}+\cdots + 
\alpha_{2\rho_2-1} \widehat{L_f^{1}h_2(x)}+ \alpha_{2\rho_2} y_{2}\\
\vdots \\
\alpha_{M1} \widehat{L_f^{\rho_M-1}h_M(x)}+\cdots + 
\alpha_{M\rho_M-1} \widehat{L_f^{1}h_M(x)}+ \alpha_{M\rho_M} y_{M}
\end{array}\right].
\eeq

Since the parameters of the dynamic functions $f_m(\cdot)$, $g_m(\cdot)$ and
$h_m(\cdot)$ are unknown, the functions $b_{m}(x_m)$ and $A_m(x_m)$ in
(\ref{66aplantflm1}) are also unknown, and so is $\xi_m(s)[y_m](t)$,
 so that the control law (\ref{6c.461}) cannot be implemented. 

\bigskip
{\bf Estimation of $\xi_m(s)[y_m](t)$}. From (\ref{66aplantflm1}),
denoting and parametrizing 
\beq
r_m(t) = \xi_m(s)[y_m](t) = b_{m}(x_m) + A_m(x_m) u_m(t) = \Theta_m^{*T}
\omega_m,
\eeq
for some unknown parameter matrix $\Theta_m^{*}$ and some known vector
signal $\omega_m(t)$, we introduce the estimate of $r_m(t)$ as
\beq
\hat{r}_m(t) = \widehat{\xi_m(s)[y_m]}(t) = \Theta_m^{T} \omega_m,
\eeq
where $\Theta_m$ is the estimate of $\Theta_m^*$. We modify the control law
(\ref{6c.461}) as
\beq
v(t) = \widehat{\xi_m(s)[y_m]}(t) - \hat{v}_y(x) = \Theta_m^{T}
\omega_m - \hat{v}_y(x),
\label{6c.462}
\eeq
which, with $\tilde{\Theta}_m = \Theta_m^* - \Theta_m$, can be expressed as
\bea
v(t) \ts = \ts - \tilde{\Theta}_m^{T} \omega_m + \Theta_m^{*T} \omega_m 
 - \hat{v}_y(x) \nn\\
\ts = \ts - \tilde{\Theta}_m^{T} \omega_m + \xi_m(s)[y_m](t)
 - \hat{v}_y(x).
\label{6c.463}
\eea
Similar to (\ref{6c.24})-(\ref{6c.33}), it follows from (\ref{6fl}) and
(\ref{6c.463}) that 
\bea
\ts \ts \left[\begin{array}{c}
e_{1}^{(\rho_1)}+\alpha_{11}e_{1}^{(\rho_1-1)}+\cdots+\alpha_{1\rho_1}e_{1}\\
e_{2}^{(\rho_2)}+\alpha_{21}e_{2}^{(\rho_2-1)}+\cdots+\alpha_{2\rho_2}e_{2}\\
\vdots \\
e_{M}^{(\rho_M)}+\alpha_{M1}e_{M}^{(\rho_M-1)}+\cdots+\alpha_{M\rho_M}e_{M}
\end{array}\right] \nn\\
\ts=\ts   \tilde{\Theta}_1^T \omega_1+ \tilde{\Theta}_2^T \omega_2 +
\tilde{\Theta}_3^T \omega_3 - \tilde{\Theta}_m^{T} \omega_m = \tilde{\Theta}^T\omega,
\label{6c.331}
\eea
where
\beq
\tilde{\Theta}=\Theta^*-\Theta
\eeq
\beq
\Theta^{T}=[\Theta_1^{T}, \Theta_2^{T}, \Theta_3^{T}, \Theta_m^T]^T
\eeq
\beq
\Theta^{*T}=[\Theta_1^{*T}, \Theta_2^{*T}, \Theta_3^{*T},
\Theta_m^{*T}]^T
\eeq
\beq
\omega=[\omega_1^T, \omega_2^T, \omega_3^T, -\omega_m^T]^T.
\eeq

The tracking error equation (\ref{6c.331}) has the same form as
(\ref{6etheta-thetas}): 
\beq
\xi_m(s)[e](t) = (\Theta^* -
\Theta(t))^T \omega(t),
\label{6etheta-thetas1} 
\eeq
based on which, a stable adaptive scheme can be designed, using the
knowledge of the measured $e(t) = y(t) - y_m(t)$ and $\omega(t)$ and
the chosen $\xi_m(s)$ formed based on the vector relative degree 
$\diag \{\rho_1, \rho_2, \ldots, \rho_M\}$ of the plant (\ref{6fsystem}).

\subsection{Adaptive Law}
We rewrite (\ref{6etheta-thetas1}) as
\beq
e(t) = W_m(s)[\tilde{\Theta}^T \omega](t),\;
\tilde{\Theta}(t) = \Theta^* - \Theta(t),
\label{62dlm1}
\eeq 
where $W_m(s) = \xi_m^{-1}(s) = \diag\{w_1(s), w_2(s), \ldots,
w_M(s)\}$ with $w_i(s) = d_i^{-1}(s)$ stable for the chosen $\xi_m(s) = \diag\{d_1(s),
d_2(s), \ldots, d_M(s)\}$. Letting  
$\tilde{\theta}_i(t) = \theta^* - \theta_i(t)$ be the $i$th column
of $\tilde{\Theta}(t) = \Theta^* - \Theta(t)$, $i = 1, 2, \ldots,
M$, we express the $i$th component of $e(t) = [e_1(t), e_2(t), \ldots,
e_M(t)]^T$ as
\beq
e_i(t) = w_{i}(s)[\tilde{\theta}_i^T \omega](t),\;i=1,2,\ldots, M.
\label{62ei}
\eeq

For $i = 1,2,\ldots, M$, introducing the auxiliary signals
\beq
\xi_{i}(t) = w_{i}(s)[\theta_i^T \omega](t) - \theta_i^T(t) \zeta_{i}(t) 
\eeq
\beq
\zeta_{i}(t) = w_{i}(s)[\omega](t),
\label{62zetaij}
\eeq
we define the estimation error signals
\beq
\epsilon_i(t) = e_i(t) + \xi_{i}(t),
\label{62epsiloni}
\eeq
and, with (\ref{62ei})-(\ref{62zetaij}), obtain 
$
\epsilon_i(t) = \tilde{\theta}_i^T(t) \zeta_{i}(t).
$

This error equation motivates the choice of the adaptive law for $\theta_i(t)$:
\beq
\dot{\theta}_i(t) =
 \frac{\Gamma_i \zeta_i(t) \epsilon_i(t)}{m_i^2(t)},
\eeq
where $\Gamma_i = \Gamma_i^T > 0$, and 
\beq
m_i(t) = \sqrt{1 + \zeta_i^T(t) \zeta_i(t)}.
\label{6mi(t)}
\eeq

\medskip
For the positive definite function $V_i
= \frac{1}{2} \tilde{\theta}_i^T \Gamma_i^{-1} \tilde{\theta}_i$, we have its
time-derivative
\beq
\dot{V}_i = - \frac{\tilde{\theta}^T \zeta_i(t) \epsilon_i(t)}{m_i^2(t)}
= - \frac{\epsilon_i^2(t)}{m_i^2(t)},
\label{6dotVi}
\eeq
from which we can derive the desired properties:

\medskip
(i) $\theta_i(t)$, $\dot{\theta}_i(t)$ and $\frac{\epsilon_i(t)}{m_i(t)}$ are 
 bounded; and 

\medskip
(ii) $\frac{\epsilon_i(t)}{m_i(t)} \in L^2$, and 
$\dot{\theta}_i(t) \in L^2$, for $i=1,2,\ldots, M$.

\medskip
\medskip
These properties are crucial for the closed-loop system signal boundedness
and asymptotic tracking performance: $\lim_{t \rightarrow \infty}
(y(t) - y_m(t)) = 0$, which can be established using the analysis
procedure for a standard adaptive feedback linearization control
scheme \cite{sb89}, \cite{si89}, \cite{wts20}, \cite{wtyz16}.

\section{Concluding Remarks}
In this paper, we have studied the extensions of the solution in
\cite{t24b} for the new model reference adaptive control problem in
which the parameters of the reference model system are unknown.

\medskip
If a dynamic system (plant) has a vector relative degree $\diag \{\rho_1,
\rho_2, \ldots, \rho_M\}$, its output $y(t)$, with 
$\xi(s) = \diag\{s^{\rho_1}, s^{\rho_2}, \ldots,
s^{\rho_M}\}$, can be expressed as
\beq
\xi(s)[y](t) = b(x) + A(x) u(t),
\label{xiybAu}
\eeq
where $b(x) \in R^M$ and $A(x) \in R^{M \times M}$ being nonsingular. 

For the linear system
case, $\xi(s)$ or $\xi(z)$ can be generalized as a lower triangular polynomial
matrix $\xi(D)$ (called the system interator matrix), and $b(x) = K_0^{*T} x(t)$
and $A(x) = K_p$ for some constant matrices $K_0^* \in R^{n \times M}$
and $K_p \in R^{M \times M}$, for (\ref{xiybAu}) with $\xi_(D)$.
 For the nonlinear system case in the
literature, the concept of $\xi(s)$ has not been generalized to a
nondiagonal matrix. The modified interactor matrix $\xi_m(s)$ is
$\xi_m(s) = \diag\{d_1(s), d_2(s), \ldots, d_M(s)\}$ with
$d_i(s)=s^{\rho_i}+\alpha_{i1}s^{\rho_i-1}+\cdots+\alpha_{i\rho_i-1} s
+\alpha_{i\rho_i}$ stable ($i=1, 2, \cdots, M$) (for the linear system
case, $\xi_m(s)$ or $\xi_m(z)$ can be a lower triangular polynomial
matrix) such that $\xi_m^{-1}(s)$ ($\xi_m^{-1}(z)$) is stable.

\medskip
With $\xi_m(s)$, the plant output $y(t)$ satisfies
\beq
\xi_m(s)[y](t) = b(x) + b_1(x) + A(x) u(t),
\eeq
for an additional vector $b_1(x) \in R^M$ ($b_1(x) = K_{01}^{*T} x(t)$
for the linear case). If the output $y_m(t)$ of the reference model
system with the state vector $x_m(t)$ and input vector $u_m(t)$ satisfies
\beq
\xi_m(s)[y_m](t) = b_m(x_m) + A_m(x_m) u_m(t),
\eeq
then, the results of this paper, as the extensions of that in \cite{t24b},
are applicable as the solutions to the output tracking problems with
unknown reference model system parameters.

For the linear continuous-time or discrete-time system case, we have
\beq
\xi_m(D)[y](t) = (K_0^* + K_{01}^*)^T x(t) + K_p u(t),
\eeq
which indicates that the solution $K_1^*$ to (\ref{mepap3}) is
$K_1^{*T} = - K_p^{-1}(K_0^* + K_{01}^*)^T$.


\begin{thebibliography}{500}
\vspace{-0.0in}
\bibitem{t24b}
Tao, G., ``Adaptive output tracking control with reference model
system  uncertainties,'' arXiv: 2406.05580 [eess.SY], 2024 (to appear
in {\it Automatica}). 

\vspace{-0.0in} 
\bibitem{t24}
Tao, G., {\it Adaptive Control under System Structure, Reference Model
  and Sensor Uncertainties}, Manuscript (to be published), 2024.
                       
\vspace{-0.0in}
\bibitem{r96}
Rugh, W. J., {\it Linear System Theory}, 2nd ed.,
 Prentice-Hall, Englewood Cliffs, NJ, 1996.

\vspace{-0.0in} 
\bibitem{t03}
Tao, G., {\it Adaptive Control Design and Analysis}, John Wiley and
Sons, New York, 2003.

\vspace{-0.0in} 
\bibitem{wtl16}
Wen, L. Y., G. Tao, and Y. Liu, ``Multivariable adaptive output
rejection of unmatched input disturbances,'' {\it
  International Journal of Adaptive Control and Signal Processing},
vol. 30, no. 8-10, pp. 1203-1227, 2016. 

\vspace{-0.0in} 
\bibitem{st21b}
Song, G. and G. Tao, ``Partial-state feedback multivariable MRAC and
reduced-order designs,'' {\it Automatica}, vol. 129, 109622, July
2021.

\vspace{-0.0in} 
\bibitem{i95}
Isidori, A., {\it Nonlinear Control Systems}, 3rd ed.,
Springer-Verlag, Berlin, 1995.

\vspace{-0.0in} 
\bibitem{sb89}
Sastry, S. and M. Bodson, {\it Adaptive Control: Stability,
Convergence, and Robustness}, Prentice-Hall, Englewood Cliffs, NJ,
1989.

\vspace{-0.0in} 
\bibitem{si89}
Sastry, S. and A. Isidori, ``Adaptive control of linearizable 
systems,'' {\it IEEE Transactions on Automatic Control}, vol. AC-34,
no. 11, pp. 1123--1131, 1989.

\vspace{-0.0in} 
\bibitem{wts20}
Wen, L. Y., G. Tao and G. Song, ``Higher-order tracking
properties of nonlinear adaptive control systems,'' {\it Systems and 
Control Letters}, vol. 145, Paper 104781, November 2020.

\vspace{-0.0in} 
\bibitem{wtyz16}
Wen, L. Y., G. Tao, H. Yang  and Y. J. Zhang, ``An adaptive
disturbance rejection control scheme for multivariable nonlinear
systems,'' {\it International Journal of Control}, vol. 89, no. 3,
pp. 594--610, March 2016.
\end{thebibliography}
\end{document}